\documentclass{emulateapj}
\usepackage{color}

\begin{document}

\shortauthors{Luhman}
\shorttitle{Gaia Survey of Lupus}

\title{
A Gaia Survey for Young Stars Associated with the Lupus Clouds\altaffilmark{1}}

\author{K. L. Luhman\altaffilmark{2,3}}

\altaffiltext{1}
{Based on observations made with the Gaia mission, the Two Micron All
Sky Survey, the Wide-field Infrared Survey Explorer, the Spitzer Space
Telescope, and the Visible and Infrared Survey Telescope for Astronomy
Hemisphere Survey.}

\altaffiltext{2}{Department of Astronomy and Astrophysics, The Pennsylvania
State University, University Park, PA 16802; kll207@psu.edu.}

\altaffiltext{3}{Center for Exoplanets and Habitable Worlds, 
The Pennsylvania State University, University Park, PA 16802, USA}

\begin{abstract}

I have used high-precision photometry and astrometry from the second data
release of the Gaia mission to perform a survey for young stars associated
with the Lupus clouds, which have distances of $\sim160$~pc and reside within
the Sco-Cen OB association.
The Gaia data have made it possible to distinguish Lupus members
from most of the stars in other groups in Sco-Cen that overlap with
the Lupus clouds, which have contaminated previous surveys.
The new catalog of candidate Lupus members should be complete for spectral
types earlier than M7 at $A_K<0.2$ within fields encompassing clouds 1--4.
I have used that catalog to characterize various aspects of the Lupus stellar
population.
For instance, the sequence of low-mass stars in Lupus is $\sim0.4$~mag
brighter than the sequence for Upper Sco, which implies an age of $\sim6$~Myr
based an adopted age of 10--12~Myr for Upper Sco and the change in luminosity
with age predicted by evolutionary models.
I also find that the initial mass function in Lupus is similar to that
in other nearby star-forming regions based on a comparison of their
distributions of spectral types.

\end{abstract}

\section{Introduction}
\label{sec:intro}

With distances of $\sim160$~pc \citep{lom08,dzi18,zuc20}, the Lupus dark clouds 
are among the nearest sites of ongoing star formation \citep{com08}.
Because of their proximity, the stars born in those clouds
are valuable for studies of the formation of stars and planets.
Candidates for stars associated with the Lupus clouds have been identified
via signatures of youth in the form of H$\alpha$ emission \citep{sch77},
X-ray emission \citep{kra97}, and mid- and far-infrared (IR)
excesses \citep{all06,all07,cha07,mer08,ben18,tei20} and via optical and
near-IR color-magnitude diagrams \citep[CMDs,][]{lop05,com09,com11,muz14}.
Many of those candidates have been observed with spectroscopy to measure their
spectral types and to help confirm their membership in the Lupus population
with spectral signatures of youth \citep{app83,hug94,kra97,wic99,com03,com13,tor06,all07,mor11,muz14,muz15,alc14,alc17,fra17}.
The resulting samples of young stars can contain both members of Lupus and
members of other populations in the Scorpius-Centaurus OB association 
\citep[Sco-Cen,][]{pm08} that overlap on the sky with the Lupus clouds.

The Gaia mission \citep{per01,deb12,gaia16b} has provided
an all-sky catalog of precise parallaxes and proper motions that can
reliably separate the stellar populations in Sco-Cen.
The second data release of Gaia (DR2) offers measurements of those parameters
for stars down to $G\sim20$ \citep{bro18}, which corresponds to
substellar masses for unobscured members of Sco-Cen.
Several studies have used the first two data releases of Gaia to 
identify candidate members of Lupus \citep{man18,mel20,tei20} as well as
Upper Scorpius, Ophiuchus, 
Upper Centaurus-Lupus (UCL), Lower Centaurus-Crux (LCC), and the V1062~Sco
moving group \citep{coo17,gol18,luh18u,ros18,wil18,can19,dam19,esp20,luh20}.
In this paper, I present a survey for members of Lupus using Gaia DR2
that improves upon previous work in terms of reducing contamination from
the other associations in Sco-Cen (Section~\ref{sec:ident}).
I have used the resulting catalog of candidate members to characterize the
age, disk fraction, initial mass function (IMF), and space velocities of the 
stellar population of Lupus clouds 1--4 and to constrain the
presence of star formation in clouds 5 and 6 (Section~\ref{sec:pop}).

\section{Identification of Candidate Members of Lupus}
\label{sec:ident}

\subsection{Defining Kinematic Membership Criteria}
\label{sec:criteria}

The data products from Gaia DR2 employed in this study include
photometry in bands at 3300--10500~\AA\ ($G$), 3300--6800~\AA\ ($G_{\rm BP}$),
and 6300-10500~\AA\ ($G_{\rm RP}$), proper motions and parallaxes
($G\lesssim20$), and radial velocities ($G\sim4$--12).
In addition, I have made use of the renormalized unit weight
error (RUWE) from \citet{lin18}, which serves as an indicator of the
goodness of fit for the astrometry. As done in Upper Sco by \citet{luh20},
I have adopted a threshold of RUWE$<$1.6 when selecting astrometry that
is likely to be reliable.

Because of projection effects, stars with similar space velocities
(e.g., members of a young association) can exhibit a broad range
of proper motions if they are distributed across a large area of sky.
As a result, populations with different space velocities can have overlapping
proper motions. To reduce such projection effects and better distinguish
the members of an association from other populations, one can analyze proper
motions in terms of their offsets from the motions expected at the celestial
coordinates and parallaxes of the stars for a characteristic space velocity
for the association (e.g., the median velocity). These proper motion offsets
have been utilized in recent studies of Gaia data in the Taurus star-forming
region and populations in Sco-Cen \citep{esp17,esp19,esp20,luh18tau,luh20}
and are used in this work as well.

\begin{figure}
\epsscale{0.9}
\plotone{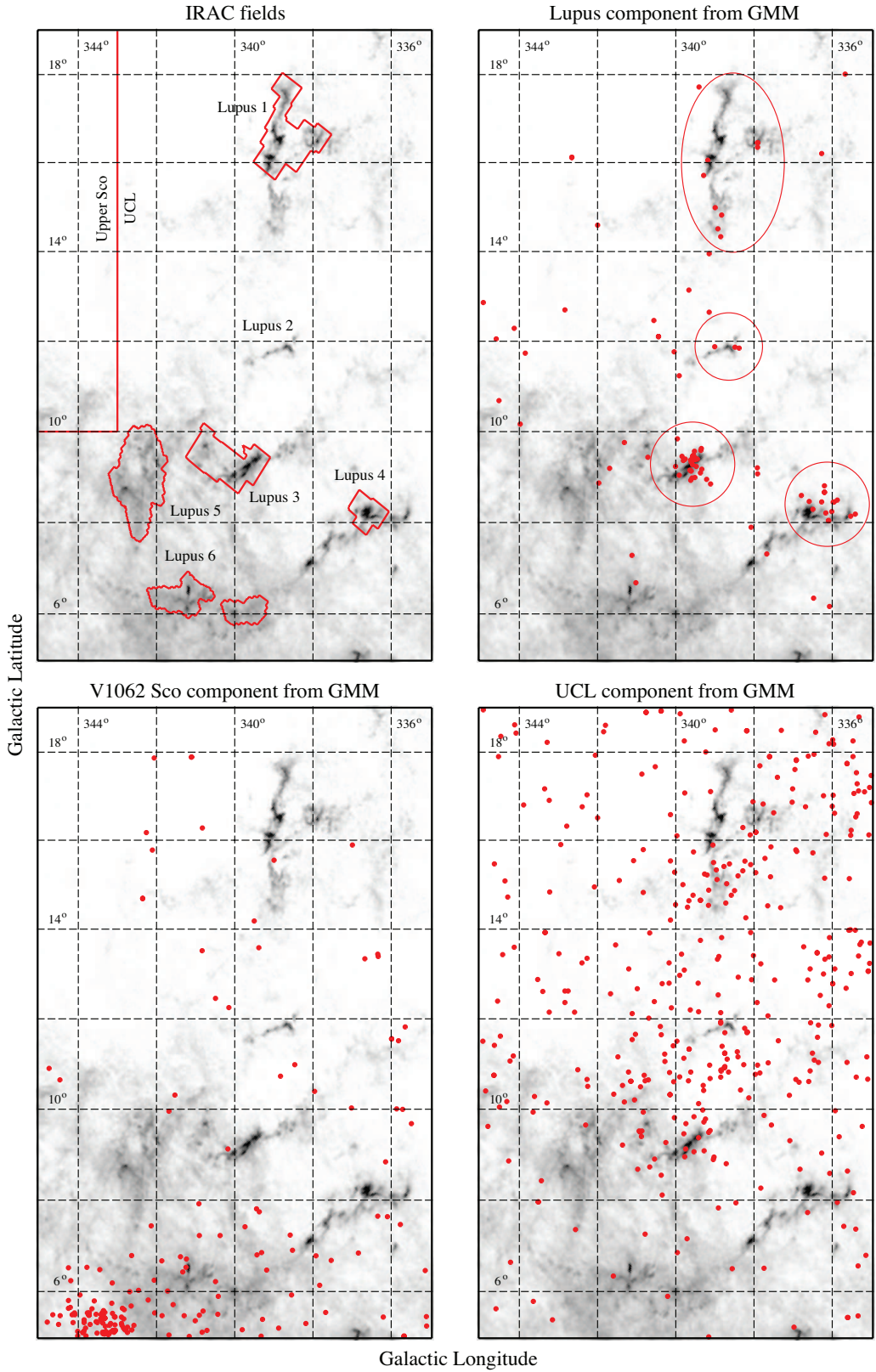}
\caption{
Maps of the fields toward the Lupus clouds that were imaged by IRAC during
the cryogenic phase of Spitzer and the three kinematic populations of
candidate young stars identified with a GMM in the middle panel of 
Figure~\ref{fig:pp}. Extinction ranging from $A_K=0.2$--2.5
is displayed with the gray scale \citep{juv16}. 
The members of the Lupus population that are within the circular and
elliptical fields encompassing clouds 1--4 are used to define the kinematics
of stars associated with those clouds (bottom panel of Figure~\ref{fig:pp}).
The map of the IRAC fields includes the boundary between Upper Sco and UCL
from \citet{dez99}.
}
\label{fig:map1}
\end{figure}

For my survey of Lupus, I have considered the area from $l=335$--$345\arcdeg$
and $b=5$--$19\arcdeg$, which is large enough to encompass clouds 1--6.
A few additional small clouds are located near or just beyond the boundary
of that field, but they do not show evidence of star formation \citep{com08}.
A map of extinction for the survey field is displayed in Figure~\ref{fig:map1}
\citep{juv16}. For reference, I have marked the boundary between Upper Sco
and UCL from \citet{dez99} and the fields that
were mapped at 3.6--8~\micron\ by the Infrared Array Camera 
\citep[IRAC;][]{faz04} on the Spitzer Space Telescope \citep{wer04}
during the facility's cryogenic phase \citep{eva03,all06,all07,mer08,spe11}.
The map does not include the IRAC observations of small areas ($\sim5\arcmin$) 
toward individual stars between the Lupus clouds or the IRAC maps of the clouds
performed during the post-cryogenic phase.
The Multiband Imaging Photometer for {\it Spitzer} \citep[MIPS;][]{rie04}
obtained images at 24~\micron\ for fields somewhat larger than the IRAC
maps in Figure~\ref{fig:map1} \citep{cha07,mer08}.

\begin{figure}
\epsscale{1.1}
\plotone{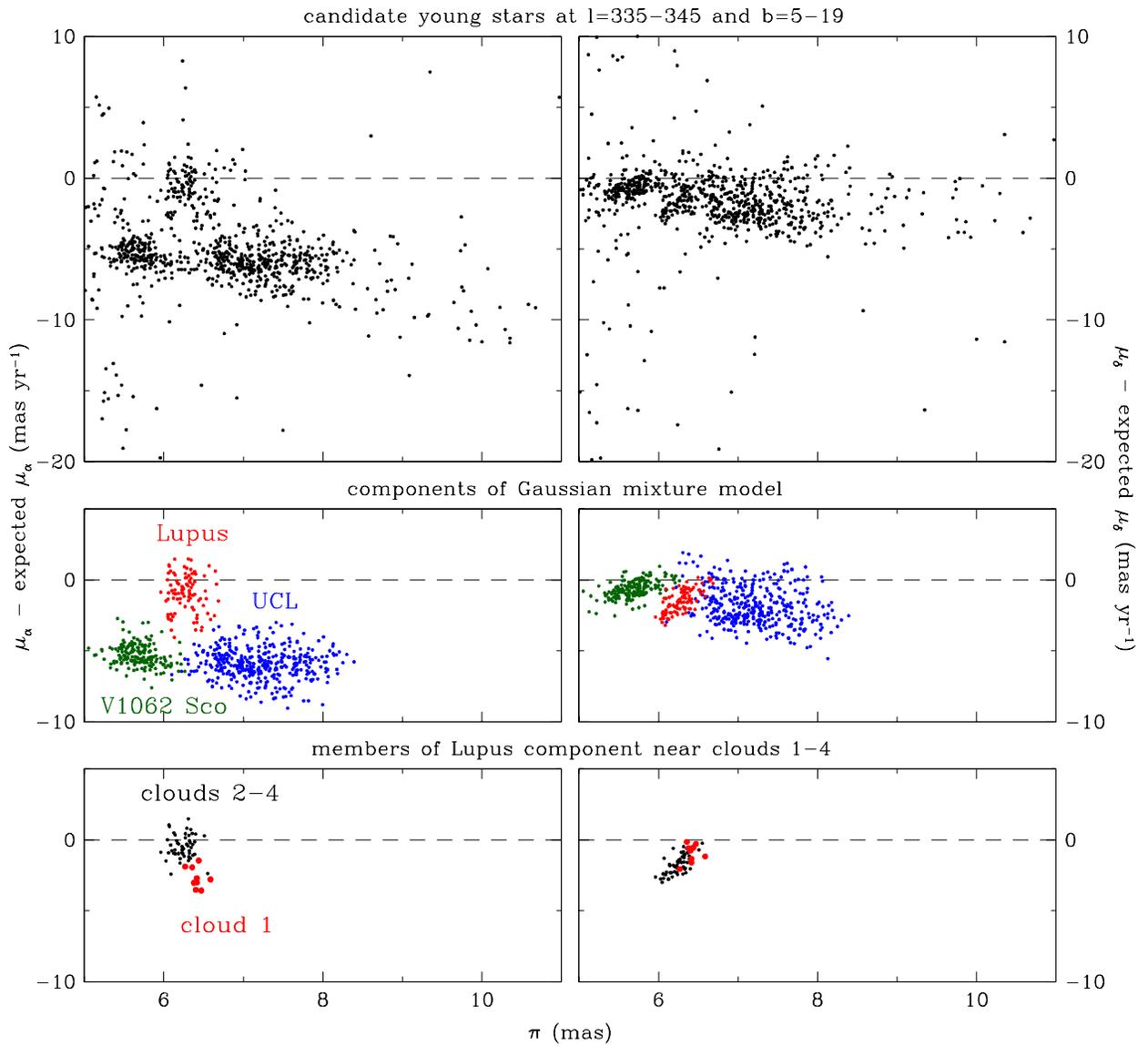}
\caption{
Proper motion offsets versus parallax for candidate young low-mass stars
($G_{\rm BP}-G_{\rm RP}$=1.4--3.4, $\sim$0.15--1~$M_\odot$) within
the boundaries of the maps in Figure~\ref{fig:map1} (top).
The offsets are computed relative to the proper motions expected
for the positions and parallaxes assuming the median space velocity of Upper
Sco members (Section~\ref{sec:criteria}).
A GMM has been used to estimate probabilities of membership
in three clustered components and a more widely scattered field component.
The stars are plotted with colors corresponding to the components to
which they most likely belong, excluding probable members of the
field component (middle). Those three components are plotted on maps in
Figure~\ref{fig:map1} and contain members of the V1062~Sco, UCL, and the
Lupus clouds. The members of the Lupus component within the circular
and elliptical fields from Figure~\ref{fig:map1} are plotted to further
refine the kinematics of stars associated with the Lupus clouds (bottom).
}
\label{fig:pp}
\end{figure}

To characterize the kinematics of the dominant populations in Sco-Cen,
\citet{luh20} used data from Gaia DR2 to construct a diagram of
$M_{G_{\rm RP}}$ versus $G_{\rm BP}-G_{\rm RP}$ for
the entire complex, selected candidate low-mass stars based on positions
above the single-star sequence for the Tuc-Hor association
\citep[45~Myr,][]{bel15}, and identified clusters in the parallaxes
and proper motion offsets of the resulting candidates.
That analysis was applied to stars with $\pi>5$~mas, $\pi/\sigma\geq20$,
RUWE$<$1.6, and $G_{\rm BP}-G_{\rm RP}$=1.4--3.4 ($\sim$K5--M5,
$\sim$0.15--1~$M_\odot$). I follow the same approach for determining
the kinematics of stars associated with the Lupus clouds, except only
the subsection of Sco-Cen selected for this survey is considered
(Figure~\ref{fig:map1}).
In the top panel of Figure~\ref{fig:pp}, I have plotted proper motion offsets
versus parallax for the candidate young low-mass stars within that field.
As done in \citet{luh20}, the proper motion offsets are computed relative
to the motions expected for a space velocity of
$U, V, W = -5, -16, -7$~km~s$^{-1}$, which approximates the median velocity
of Upper Sco. That velocity is also similar to the median value for
members of Lupus (Section~\ref{sec:uvw}).
Three concentrations of stars are evident in the proper motion offsets
and parallaxes in Figure~\ref{fig:pp}.
The middle population in parallax contains stars grouped near Lupus clouds
1--4 in both celestial coordinates and parallax \citep{lom08,dzi18,zuc20}.
It also coincides with the kinematics of Upper Sco and likely
includes members of that association, as discussed in Section~\ref{sec:ages}.
The other two populations correspond to the V1062~Sco group and UCL.

\citet{luh20} estimated probabilities of membership in three Sco-Cen
populations (V1062~Sco, Upper Sco/Ophiuchus/Lupus, UCL/LCC) and a field
population by applying a Gaussian mixture model (GMM)
to the proper motion offsets and parallaxes for candidate young low-mass stars
across the entirety of Sco-Cen. I have performed the same modeling on
the data in the top panel of Figure~\ref{fig:pp} for my survey field
containing the Lupus clouds, which is done using the {\tt mclust} package in
R \citep{Rcore13,mclust}. The resulting membership assignments to three
components corresponding to Lupus (and Upper Sco), V1062~Sco, and UCL are
indicated by the colors of the points in the middle panel of
Figure~\ref{fig:pp}. The stars that are more likely
to reside in the field component have been omitted.
Most stars have high probabilities of membership in a single component.
For instance, 87 of the 94 stars in the Lupus component have 
membership probabilities of $>$80\%. The spatial distributions
of the members of the three GMM components are shown on maps in
Figure~\ref{fig:map1}. Most members of the V1062~Sco component are
concentrated near that star ($l\sim343.6\arcdeg$, $b\sim5.2\arcdeg$) 
while the UCL component is widely scattered across the survey field.
A majority of the stars assigned to the Lupus component are grouped near
clouds 1--4. These maps illustrate that young stars with the kinematics
of V1062~Sco and UCL represent a potential source of contamination
in surveys for Lupus members that are based only on evidence of youth.

As shown in Figure~\ref{fig:map1}, the GMM component for Lupus contains
stars grouped near clouds 1--4 and a smaller number of stars that are widely
scattered across the survey field. In Section~\ref{sec:ages}, I find that
the latter are likely to be members of Upper Sco, so I have attempted to
refine the characterization of the kinematics of Lupus members by considering
only the stars near the clouds. To do that, I have defined circular boundaries
that encompass the groups of stars near clouds 2--4 and an elliptical
boundary that is large enough to contain cloud 1, which are marked in
Figure~\ref{fig:map1}. In this study, stars within and outside of
those boundaries are referred to as ``on cloud" and ``off cloud", respectively.
The bottom panel of Figure~\ref{fig:pp} shows the
proper motion offsets and parallaxes of the 61 on-cloud members of the GMM
component for Lupus. The distributions of those
parameters are tighter with the omission of the off-cloud stars.
The kinematics of the stars near cloud 1 exhibit small offsets relative
to the stars near clouds 2--4. \citet{dzi18} found a similar shift
when comparing the average proper motions of young stars near the clouds.
Among the 61 on-cloud stars, 52 have been observed previously with
spectroscopy, all of which have evidence of ages young enough for membership
in Sco-Cen ($\lesssim20$~Myr). 
In the next section, I will use the proper motion offsets and parallaxes of
the 61 on-cloud stars from the GMM component to define criteria
for a search for candidate members of Lupus.

\subsection{Applying Kinematic Membership Criteria}
\label{sec:apply}

\begin{figure}
\epsscale{1.2}
\plotone{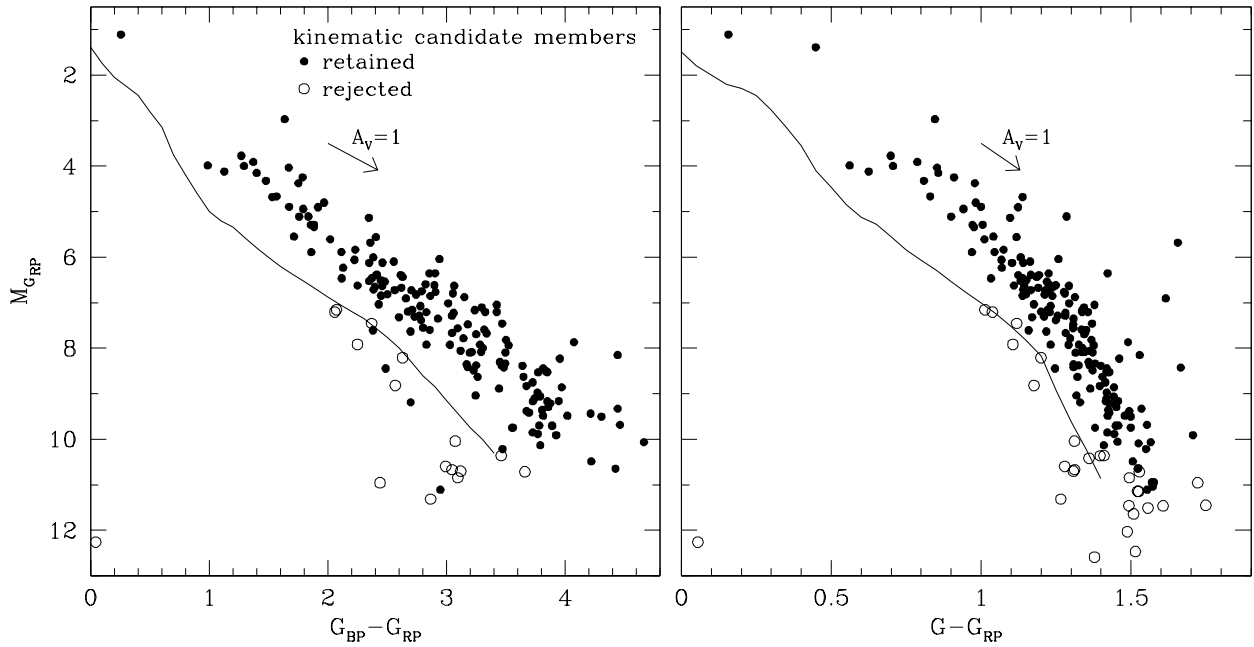}
\caption{
$M_{G_{\rm RP}}$ versus $G_{\rm BP}-G_{\rm RP}$ and $G-G_{\rm RP}$ for
candidate members of Lupus selected via their kinematics from Gaia DR2,
which consist of stars that have positions within the boundaries of the maps in 
Figure~\ref{fig:map1}, $\pi/\sigma\geq10$, and parallaxes and proper motion
offsets overlapping with those of the stars in the bottom panel of
Figure~\ref{fig:pp}.
Each diagram includes a fit to the single-star sequence for the Tuc-Hor
association \citep[45~Myr,][]{bel15} (solid line).
Stars that appear above the Tuc-Hor sequence in either diagram or exhibit
mid-IR excess emission from a circumstellar disk are retained as candidate
members of Lupus (filled circles) while the remaining stars are rejected
(open circles). The reddening vectors are based on the extinction curve
from \citet{sch16}.
}
\label{fig:cmd1}
\end{figure}

I have searched for members of Lupus by selecting
stars from Gaia DR2 that have positions between $l=335$--$345\arcdeg$ and
$b=5$--$19\arcdeg$ (the boundaries of the maps in Figure~\ref{fig:map1}),
$\pi/\sigma\geq10$, and parallaxes and proper motion offsets that overlap
at $1\sigma$ with those of the 61 on-cloud stars from the GMM component
for Lupus in the bottom panel of Figure~\ref{fig:pp}.
The resulting sample contains 206 stars.
To further refine these candidates, I have plotted them on diagrams
of $M_{G_{\rm RP}}$ versus $G_{\rm BP}-G_{\rm RP}$
and $M_{G_{\rm RP}}$ versus $G-G_{\rm RP}$ in Figure~\ref{fig:cmd1},
which includes a fit to the single-star sequence for the Tuc-Hor association
\citep[45~Myr,][]{bel15}.
Data with errors greater than 0.1~mag have been excluded.
Nineteen stars appear in the diagram with $G-G_{\rm RP}$ but not in the
diagram with $G_{\rm BP}-G_{\rm RP}$.
I have rejected the candidates that are likely too old to be members of Lupus
based on positions in those diagrams near or below the sequence for Tuc-Hor.
Stars with circumstellar disks can appear unusually faint for their color
in CMDs due to short-wavelength excess emission
from accretion or scattered light from an occulting disk, so all
candidates with mid-IR excess emission indicative of disks have been retained.
A total of 28 candidates were rejected based on the CMDs.
Seven of the 206 candidates do not appear in either of the CMDs because they
lack photometry in $G_{\rm BP}$ and $G_{\rm RP}$. Six of these stars are
possible companions to brighter stars that appear to be young in CMDs
and the remaining candidate exhibits evidence of youth in the form of
mid-IR excess emission, so these seven candidates are retained.
After rejection of 28 candidates, there remain 178 candidates for kinematic
members of Lupus.

It is likely that a small number of Lupus members were not selected by my
kinematic criteria due to poor astrometric fits or large errors.
Some of these members can identified if they are companions to members
that satisfied the kinematic criteria. I searched Gaia DR2
for stars that are located within $10\arcsec$ of any of the 178 kinematic
candidates, have positions on CMDs that are consistent with membership,
and have parallaxes and proper motions within the range of values in Sco-Cen.
Nine candidate members were found in that manner, some
of which have large values of RUWE, indicating poor astrometric fits,
or have large parallaxes errors ($\pi/\sigma<10$).
These candidates consist of RX~J1529.7-3628A, RX~J1609.9-3923B,
RX~J1539.7-3450B, and sources 5997033290348155392, 5997083524280365056,
5994795990331624704, 6018569458962613888, 6035794025846997120,
and 6007849461103724672 from Gaia DR2. 

As an additional check for Lupus members that were missed by the kinematic
criteria due to erroneous astrometry, I selected stars from Gaia DR2 that
have positions in the on-cloud fields, $\pi/\sigma\geq10$,
CMD ages that are consistent with membership, and kinematics
that do not satisfy my criteria for Lupus membership. For the resulting
stars that exhibited signatures of youth from 
previous spectroscopy, I inspected the available astrometry for evidence
that the Gaia DR2 measurements might be unreliable.
Four stars were identified through this process, consisting of
Gaia DR2 6011522757643072384, HD~142527, and Sz~108A/B.
The first star is a $16\arcsec$ candidate companion to GQ~Lup \citep{alc20}
and may have a poor astrometric fit (RUWE=2.25).
HD~142527 is a well-studied Herbig Ae/Be star \citep{wae96} that has a
low-mass stellar companion at a separation of $<0\farcs1$ \citep{bil12,clo14}.
It failed the kinematic criteria due to its proper motion in declination
from DR2 ($-24.460\pm0.052$~mas~yr$^{-1}$), 
but it satisfies the criteria if the significantly different measurement
from DR1 is adopted ($-26.336\pm0.065$~mas~yr$^{-1}$).
Sz~108A and B are known young stars \citep{app83,hug94,com03}
that appear to comprise a binary system given their small separation
($4\arcsec$) but that exhibit different parallaxes and proper motions
from Gaia DR2 (e.g., 6.61$\pm$0.05~mas vs. 5.92$\pm$0.12~mas).
The individual stars do not satisfy the membership criteria in both
the parallax and proper motion offsets, but the average values for the pair
are consistent with membership (i.e., the kinematics of the stars are on
opposite sides of the Lupus criteria). I have included these four stars in the
sample of candidates.

The 191 candidate members of Lupus from the preceding analysis are presented
in Table~\ref{tab:mem1}.
The catalog includes previous measurements of spectral types and radial
velocities, astrometry and photometry from Gaia DR2,
$UVW$ space velocities computed from the radial velocities and
Gaia astrometry (Section~\ref{sec:uvw}), near-IR photometry from the
Point Source Catalog of the Two Micron All Sky Survey \citep[2MASS,][]{skr06}
and other sources, mid-IR photometry from the AllWISE Source Catalog of
the Wide-field Infrared Survey Explorer \citep[WISE,][]{wri10},
flags for the presence of mid-IR excess emission and
disk classifications (Section~\ref{sec:disks}), and a flag
indicating whether the celestial coordinates are within the circular and
elliptical boundaries toward clouds 1--4 in Figure~\ref{fig:map1}
(i.e., on cloud).
For each pair of Gaia sources that is unresolved in 2MASS or WISE, the
data from the latter surveys have been assigned to the component that is
brighter in the $G$ band with the exception of Sz~108~A/B, where the
dominant component varies among the WISE bands.
Among the 191 candidates in Table~\ref{tab:mem1}, 121 are on cloud and
106 have been observed previously with spectroscopy, all of which show
evidence of youth ($\lesssim20$~Myr) from Li~I, gravity-sensitive absorption
lines, or IR excess emission.

In addition to the Lupus candidates with Gaia kinematics in
Table~\ref{tab:mem1}, I have compiled candidates that
lack parallaxes and proper motions from Gaia and that are located
within the circular and elliptical boundaries toward clouds 1--4 in
Figure~\ref{fig:map1}.
I have retrieved sources from the 2MASS Point Source Catalog within those
fields and have rejected those that have Gaia kinematics or CMDs that
are inconsistent with membership and those that are not young stars based
on other available data (e.g., resolved galaxies).
The number of remaining stars is low down to $H\sim14$ and increases
rapidly at fainter magnitudes.
I have adopted the 18 unrejected 2MASS sources at $H<14$ as candidates.
Among those candidates, 10 have previous spectral classifications,
12 exhibit evidence of youth from spectroscopy or mid-IR
excess emission, and 13 have counterparts in Gaia DR2.
All of the latter stars have large values of RUWE that indicate poor
astrometric fits, which explains the absence of accurate measurements of
parallaxes and proper motions.
I also have examined available constraints on membership for
candidates for disk-bearing stars that have been previously identified in
mid- and far-IR imaging \citep[e.g.,][]{mer08,ben18} and that lack Gaia data.
Five of those stars are adopted as candidate members,
which include the known protostars IRAS~15398$-$3359 and Lupus~3~MMS.
The combined sample of 23 on-cloud candidates lacking Gaia kinematics is
presented in Table~\ref{tab:mem2}, which contains the same data as in
Table~\ref{tab:mem1} with the exception of the kinematic data.
The $0\farcs7$ companion GQ~Lup~B is an additional probable member of
Lupus \citep{neu05}, but it is omitted from Table~\ref{tab:mem2}
because it lacks most of the tabulated measurements.

\subsection{Comparison to Previous Surveys}
\label{sec:compare}

Several previous studies have presented candidate members of Lupus.
For the candidates with measurements of parallaxes and proper motions
from Gaia, I have examined whether they appear in my catalog of candidates.
If not, I have used the Gaia data to assess whether the candidates are
field stars or members of other populations in Sco-Cen.

\citet{kra97} identified young stars in a large area surrounding the
Lupus clouds through optical spectroscopy of X-ray sources, 74 of which
are located within the field considered in this work ($l=335$--$345\arcdeg$,
$b=5$--$19\arcdeg$) and have parallaxes and proper motions from Gaia DR2.
Based on the Gaia kinematics, I have classified 14 as candidate Lupus
members, 17 as field stars, and 43 as candidate members of other
populations in Sco-Cen (primarily UCL).
It is not surprising that a wide-field survey for young stars would be
dominated by the latter given the results in Figure~\ref{fig:map1}.

\citet{all07} selected 19 candidates for disk-bearing stars from
Spitzer images of cloud~3, 10 of which were confirmed as young stars
through spectroscopy in that work and subsequent studies
\citep{mor11,muz15,alc14,alc17}.
Based on Gaia parallaxes and proper motions, I have classified the confirmed
young stars as six candidate members of Lupus, three candidate members of
other Sco-Cen groups, and one field star, although the errors for two
stars are fairly large ($\sim15$\% in parallax).
Gaia data are available for two of the nine candidates from \citet{all07}
that lack spectra, which I classify as a background star and a
candidate member of V1062~Sco or UCL. Four additional
candidates that lack spectra are galaxies \citep[][VISTA]{com11,com13}.

\citet{mer08} presented a large sample of candidate young stars
from Spitzer images of clouds 1, 3, and 4.
Among the 126 candidates with Gaia kinematics,
I have classified 56 as Lupus candidates, 22 as candidate members of other
groups in Sco-Cen, and 48 as unrelated to Sco-Cen. 
This level of contamination by other Sco-Cen groups
is consistent with the surface densities of their
kinematic members \citep[][Figure~\ref{fig:map1}]{dam19}.

\citet{com09} identified candidate members of Lupus using optical images
of clouds 1, 3, and 4 that encompassed the Spitzer fields from \citet{mer08}. 
Based on my kinematic analysis, the 140 stars with Gaia parallaxes
and proper motions consist of 19 candidate Lupus members, 19 candidate members
of the remainder of Sco-Cen, and 102 objects that are unrelated to Sco-Cen.
\citet{com13} also found that many of the candidates
are background stars based on spectroscopy.

\citet{man18} used kinematic data from Gaia DR2 to examine the membership
of candidate members of clouds 5 and 6 from \citet{spe11}.
They found that five candidates exhibit parallaxes and proper motions
that are suggestive of membership in Lupus while the remaining candidates
with Gaia data are background stars.
One of those five candidates satisfies my kinematic criteria for Lupus
membership (Gaia DR2 6021805356032645504), one has a discrepant parallax
for Lupus (Gaia DR2 6021745462701109376, $\pi=8.91\pm0.19$~mas),
and the remaining three candidates are just beyond my kinematic thresholds
for selection for Lupus (Gaia DR2 6021420630046381440, 6021662385163162240,
6021662385163163648).

\citet{dam19} used Gaia DR2 to identify candidate members of various
populations in Sco-Cen, including cloud 3 in Lupus. For the 69 stars assigned
to that cloud, I have classified 61 as Lupus candidates, five as candidate
members of other Sco-Cen groups, and three as unrelated to Sco-Cen.

\citet{mel20} assumed Lupus membership for 154 previously identified
candidate members that had Gaia parallaxes and proper motions and
searched Gaia DR2 for additional candidates that exhibited similar kinematics.
I find that 65 of the stars adopted as members have kinematics inconsistent
with membership in Lupus, most of which are likely members of other
populations in Sco-Cen.
Because the adopted Lupus members were contaminated by non-members,
the candidates identified in that search of Gaia DR2 were contaminated as well.
Among the 431 candidates, 50 are classified as Lupus candidates in this work
while the remaining stars are classified as non-members.

\citet{tei20} selected candidate disk-bearing stars 
in a large field encompassing the Lupus clouds ($l=330$--$349\arcdeg$,
$b=1.6$--$27.6\arcdeg$) using mid-IR photometry from WISE.
They checked whether the candidates were likely to be members of
Lupus, UCL, V1062~Sco, or Upper Sco by comparing their proper motions and
parallaxes to those of previously proposed members.
Lupus and UCL appeared to have indistinguishable proper motions,
so candidates were classified as members of either of those
two populations.
However, I find that Lupus and UCL do exhibit distinct kinematics
when analyzed in terms of proper motions that have been corrected
for projection effects (Figure~\ref{fig:pp}) and when the previous samples
in Lupus members are vetted for contamination from UCL and other groups in
Sco-Cen. \citet{tei20} presented 60 candidate members of Lupus or UCL, three
of which are among my kinematic Lupus candidates in Table~\ref{tab:mem1}.
I classify the remaining 57 candidates as field stars or members
of other groups in Sco-Cen based on their parallaxes and proper motion offsets.

\section{Properties of the Lupus Stellar Population}
\label{sec:pop}

\subsection{Spectral Types and Extinctions}
\label{sec:spt}

I seek to characterize the stellar population associated
with the Lupus clouds in terms of its age, disk fraction, IMF, and
space velocities using the candidate members from Tables~\ref{tab:mem1}
and Table~\ref{tab:mem2}. Some of this analysis requires estimates
of extinctions and spectral types.
For stars with previously measured spectral types, I have estimated
extinctions from color excesses in $G_{\rm RP}-J$ or $J-H$
(in order of preference) relative to the typical intrinsic colors
of young stars from \citet{luh20}.
The reddenings are converted to extinctions in $K_s$ using
relations from \citet{ind05} and \citet{sch16} and
$E(G_{\rm RP}-J)/E(J-H)\approx2.4$, where the latter is
based on reddened members of Upper Sco and Ophiuchus \citep{esp20,luh20}.

\begin{figure}
\epsscale{1.2}
\plotone{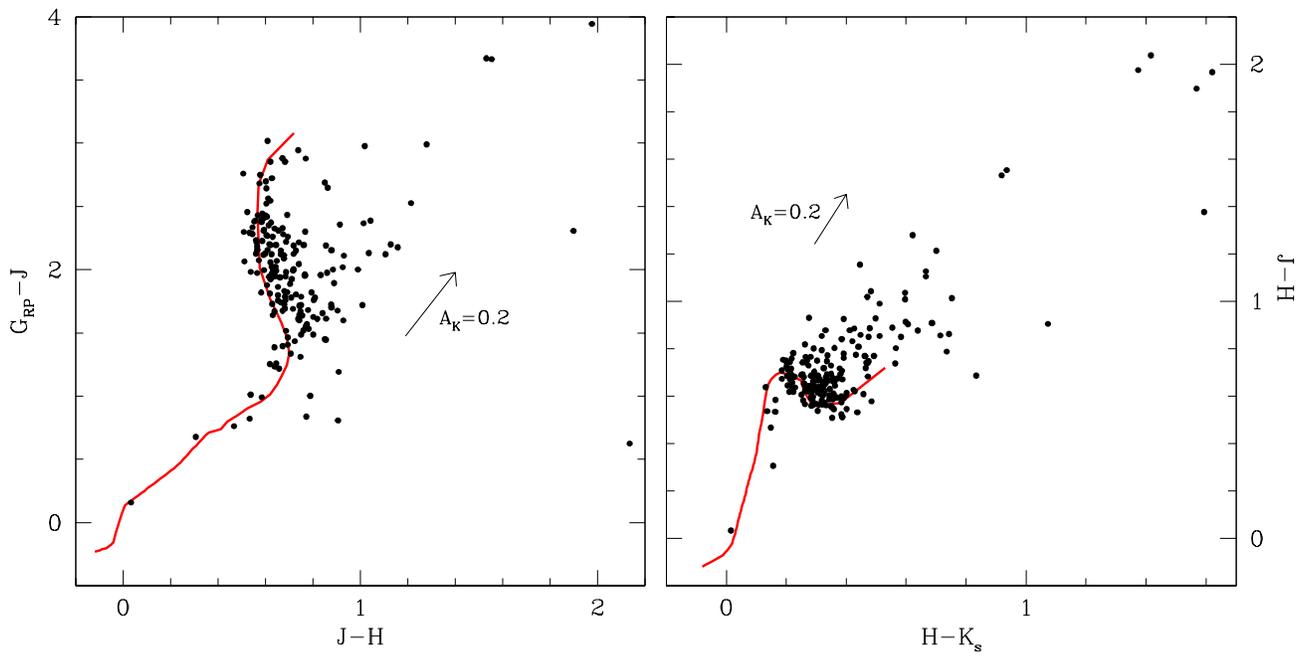}
\caption{
$G_{\rm RP}-J$ versus $J-H$ and $J-H$ versus $H-K_s$ for candidate members
of Lupus from Tables~\ref{tab:mem1} and \ref{tab:mem2} (filled circles).
The intrinsic colors of young stars from B0--M9 are indicated
\citep[red solid lines,][]{luh20}.
}
\label{fig:cc}
\end{figure}

For stars that lack spectral classifications, I have estimated
spectral types and extinctions by dereddening their observed colors
to the sequence of intrinsic colors of young stars in diagrams of 
$G_{\rm RP}-J$ versus $J-H$ or $J-H$ versus $H-K_s$, which are 
shown in Figure~\ref{fig:cc} for the Lupus candidates.
Since the sequence of intrinsic colors in $G_{\rm RP}-J$ versus $J-H$
is largely parallel to the reddening vector at earlier types, spectral
types and extinctions are estimated from that diagram only for 
$G_{\rm RP}-J>1.3$ (M types). With the exception of a few stars with $J-H$
excesses from disks, the stars at $G_{\rm RP}-J<1.3$ are quite close to
the sequence of intrinsic colors, so they are assumed to have no extinction
and their spectral types are estimated from the observed values of
$G_{\rm RP}-J$. For the few stars that lack $G_{\rm RP}$ but have $JHK_s$,
spectral types and extinctions are estimated from $J-H$ versus $H-K_s$
if photometry at longer wavelengths does not show excess emission from
circumstellar dust (Section~\ref{sec:disks}).
For companions that have only Gaia photometry, the extinctions of their
primary stars are adopted, and their spectral types are estimated
from extinction-corrected $G_{\rm BP}-G_{\rm RP}$ if available.
If a companion only has $G$ photometry, a spectral type is derived
from the median relation between extinction-corrected $G$ and spectral type
among the candidates that have spectral classifications.
I have not attempted to estimate spectral types and extinctions for 
the small number of candidates with $J-H>1.2$ ($A_K\gtrsim0.5$ for M stars)
since the upcoming analysis that involves those parameters
(Sections~\ref{sec:ages} and \ref{sec:imf}) excludes stars with high
extinctions.

\subsection{Stellar Ages}
\label{sec:ages}

The age of a young stellar population can be constrained via its sequence
of low-mass stars in the Hertzsprung-Russell (H-R) diagram.
Photometry in a single band and either spectral types or colors can serve as
substitutes for bolometric luminosities and effective temperatures,
respectively.
For instance, in a recent study of Ophiuchus, \citet{esp20} analyzed
ages using an IR photometric band and spectral types since optical photometry
and colors would be significantly affected by the high extinction among many
of the members of that region. For Lupus, most of the candidate members
have only modest extinction (Figure~\ref{fig:cc}) and many of them lack
spectral classifications, so the high-precision optical photometry from
Gaia is the best option for both axes of the H-R diagram (i.e., a CMD). 
The same approach was taken by \citet{luh20} when comparing the ages of
Upper Sco, UCL, LCC, and V1062~Sco, all of which have low extinction.

\begin{figure}
\epsscale{1.2}
\plotone{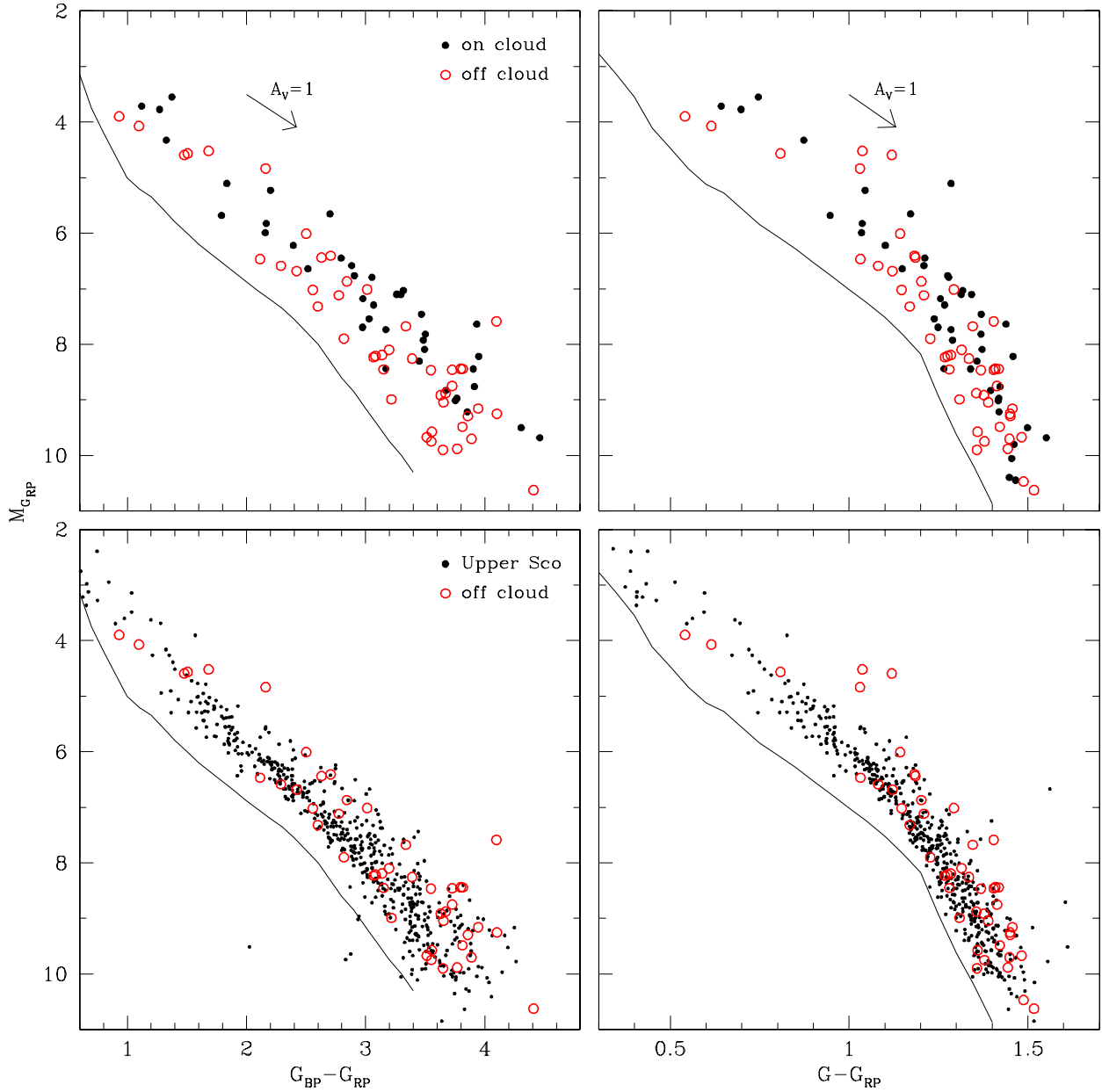}
\caption{
Extinction-corrected $M_{G_{\rm RP}}$ versus $G_{\rm BP}-G_{\rm RP}$ and
$G-G_{\rm RP}$ for the candidate members of Lupus from
Table~\ref{tab:mem1} that have precise parallaxes ($\pi/\sigma\geq20$) and
low extinctions ($A_K<0.2$) and do not have full disks (top).
Candidates within the circular and elliptical fields in Figure~\ref{fig:map1}
are plotted with filled circles and candidates outside of those fields
are plotted with open circles.
The off-cloud candidates are shown with members of Upper Sco that have
$\pi/\sigma\geq20$ and $A_K<0.1$ and that lack disks (bottom).
Each diagram includes a fit to the single-star sequence for the Tuc-Hor
association \citep[45~Myr,][]{bel15} (solid line).
}
\label{fig:cmd2}
\end{figure}

In the top row of Figure~\ref{fig:cmd2}, I have plotted diagrams of
$M_{G_{\rm RP}}$ versus $G_{\rm BP}-G_{\rm RP}$ and $M_{G_{\rm RP}}$ versus
$G-G_{\rm RP}$ for Lupus candidates from Table~\ref{tab:mem1} that have
$\pi/\sigma\geq20$ and $A_K<0.2$ and that lack full disks
(Section~\ref{sec:disks}). Stars with full disks have been omitted because 
$G_{\rm BP}$ is susceptible to accretion-related emission at short optical
wavelengths.
The photometry in the CMDs has been corrected for the extinctions estimated
in the previous section. The on- and off-cloud candidates are plotted with 
different symbols in Figure~\ref{fig:cmd2}, which shows that the off-cloud
stars are systematically fainter at a given color, indicating older ages.
The off-cloud stars also exhibit a lower disk fraction
(Section~\ref{sec:disks}), which is consistent with older ages.

In the bottom row of Figure~\ref{fig:cmd2}.
the off-cloud candidates are compared to diskless members of Upper Sco
that have $\pi/\sigma\geq20$ and $A_K<0.1$ \citep{luh20}.
The lower envelopes of the sequences for those two samples are roughly
aligned in absolute magnitude, which suggests that the oldest off-cloud stars
are coeval with Upper Sco \citep[10--12~Myr,][]{pec12,pec16,luh20}.
For the colors corresponding to K4--M5 ($G_{\rm BP}-G_{\rm RP}\sim1.4$--3.3,
$G-G_{\rm RP}\sim0.7$--1.3), the sequences for the on-cloud Lupus candidates
in the two CMDs are $\sim0.4$~mag brighter than the sequences for Upper Sco.
If the latter has an age of 10--12~Myr, that difference implies an age
of 5.2--6.5~Myr for Lupus based on non-magnetic evolutionary models
\citep{bar15,cho16,dot16} and new versions of the magnetic models from
\citet{fei16} provided by G. Feiden (private communication).
The sample of on-cloud candidates in Figure~\ref{fig:cmd2}
contains 30 stars near cloud 3 and only 1--9 stars near each of the
other clouds, so the numbers of stars are too small for a comparison of ages
among the clouds.

Given the difference in mean ages of the on- and off-cloud Lupus candidates,
most of the off-cloud candidates are probably not associated with the Lupus
clouds. Instead, they are likely members of a low-density,
distributed component of Upper Sco that extends into the survey field
(see the nominal boundary between Upper Sco and UCL in Fig.~\ref{fig:map1}),
which is supported by the coevality of the oldest off-cloud candidates with
Upper Sco and the fact that the Lupus candidates have parallaxes and proper
motion offsets within the ranges of values exhibited by Upper Sco \citep{luh20}.
Therefore, the best sample of Lupus candidates consists of the on-cloud
stars in Table~\ref{tab:mem1} and the candidates in Table~\ref{tab:mem2}
(all of which are on cloud). The surface density of off-cloud candidates
implies that the on-cloud sample could contain $\sim10$ stars that are
members of Upper Sco rather than Lupus.
Meanwhile, it is possible that a few of the off-cloud candidates
are associated with the Lupus clouds.

\subsection{Disk Fraction}
\label{sec:disks}

I have used mid-IR photometry from WISE and the Spitzer Space Telescope to
check for evidence of circumstellar disks around the candidate members of Lupus.
WISE obtained images in bands centered at 3.5, 4.5, 12, and 22~$\mu$m, which
are denoted as W1, W2, W3, and W4, respectively.
Spitzer observed primarily in bands at 3.6, 4.5, 5.8, 8.0 and 24~\micron, which
are denoted as [3.6], [4.5], [5.8], [8.0], and [24], respectively.
As mentioned in Section~\ref{sec:criteria},
Spitzer has imaged areas encompassing most of the Lupus clouds
(Figure~\ref{fig:map1}) as well as small fields toward individual stars
between the clouds. As an all-sky survey, WISE provides images that cover all
of the Lupus candidates, but it offers lower sensitivity and angular resolution
than Spitzer. 

As done in previous disk surveys by \citet{luh12} and \citet{esp14,esp18},
I have used the extinction-corrected colors between $K_s$ and six
bands from WISE and Spitzer (W2, W3, W4, [4.5], [8.0], [24]) 
to detect excess emission from disks.
For each extinction-corrected color, I have subtracted the typical
intrinsic color for the spectral type in question \citep{luh20}.
The resulting color excesses are used to determine whether significant disk
emission is present and to classify the evolutionary stages of any detected
disks \citep{esp14,esp18}. The adopted disk classes consist of the following
\citep{ken05,rie05,her07,luh10,esp12}:
full (optically thick with no large holes),
transitional (optically thick with a large hole),
evolved (optically thin with no large hole),
evolved transitional (optically thin with a large hole),
and debris disk (second generation dust from planetesimal collisions).
For reference, young stars and their circumstellar material can be
classified in the following manner
\citep{lw84,lad87,and93,gre94}:
classes~0 and I (protostar with an infalling envelope and a full disk),
class~II (star with a primordial disk but no envelope), and class~III
(star that no longer has a primordial disk but that can have a debris disk).
The class~II disks include full, transitional, evolved, and evolved
transitional, although the latter are sometimes counted as class~III
when calculating disk fractions since they are indistinguishable from
debris disks in mid-IR data.

\begin{figure}
\epsscale{1.4}
\plotone{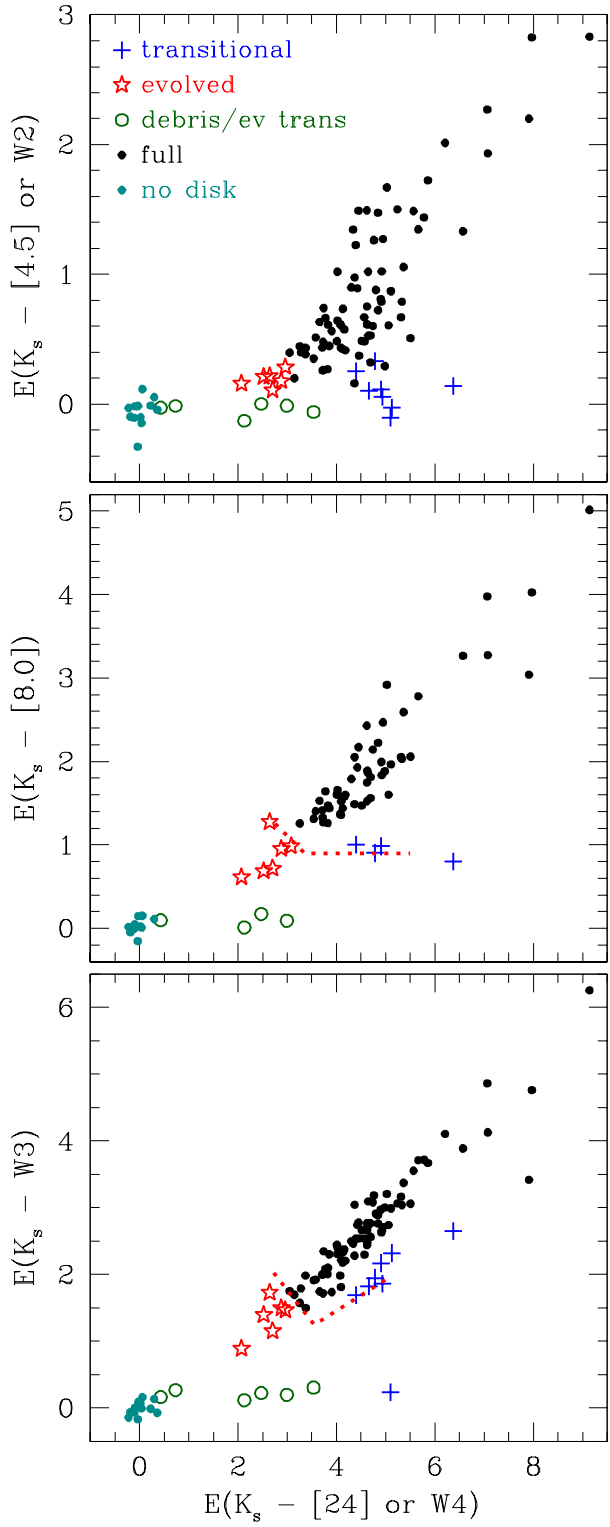}
\caption{
Extinction-corrected IR color excesses for candidate members of Lupus
from Tables~\ref{tab:mem1} and \ref{tab:mem2}.
Data at [4.5] and [24] are shown when available.  Otherwise, measurements
at similar wavelengths from WISE are used (W2 and W4).
The bottom two diagrams include boundaries that are used to distinguish full
disks from disks in more advanced stages of evolution (dotted lines).}
\label{fig:ex}
\end{figure}

The color excesses for the Lupus candidates are plotted in Figure~\ref{fig:ex}.
Tables~\ref{tab:mem1} and \ref{tab:mem2} include flags that indicate whether
excesses are detected in W2, W3, W4, [4.5], [8.0], and [24]
and the classifications of the disks.
A few of the Lupus candidates are listed in Tables~\ref{tab:mem1} and
\ref{tab:mem2} as class~0 or class~I based on previous work.
Candidates that do not show excesses in their mid-IR data are labeled as 
class~III. Among the 214 candidates in Tables~\ref{tab:mem1} and \ref{tab:mem2},
17 lack mid-IR photometry because they are below the detection limits
of the available imaging or are unresolved from brighter companions.

Most of the disks identified in my analysis have been found in previous 
studies \citep{eva03,pad06,cie07,mer08,mer10,wah10,rom12,bus15,pec16,ben18}.
The newly detected disks include three transitional disks
(Sz~127, Gaia DR2 6013489268547136768, Gaia DR2 6035036466655228672) and two
evolved disks (Gaia DR2 5995219680274444672, Gaia DR2 6014623517871055488).
Sz~127 has been classified as M5 \citep{hug94} while the remaining four
stars have not been observed with spectroscopy.
Gaia DR2 6013489268547136768 and 6035036466655228672 are far from the Lupus clouds,
so they are probably members of Upper Sco (Section~\ref{sec:ages}).

\begin{figure}
\epsscale{1.1}
\plotone{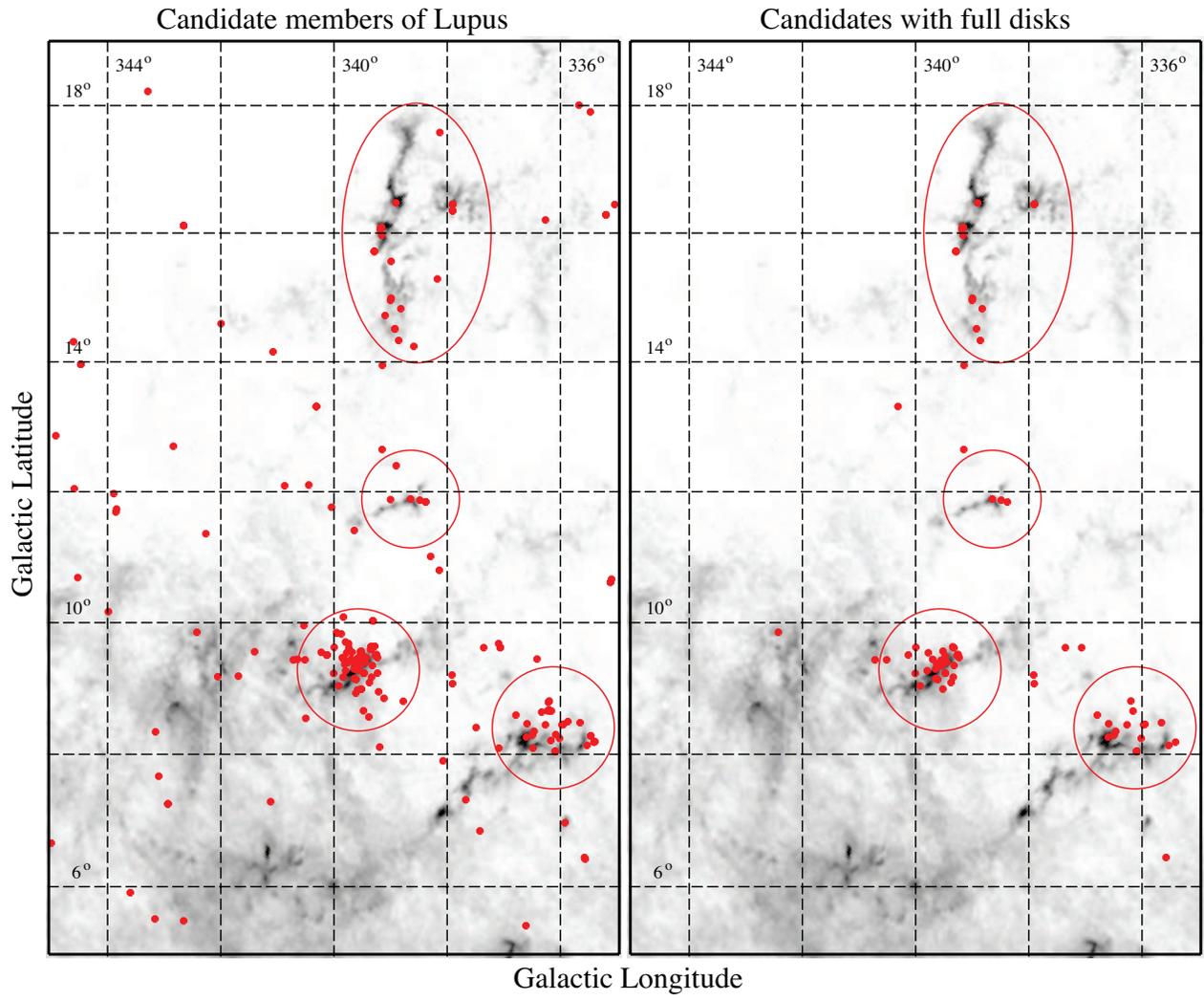}
\caption{
Map of the candidate members of Lupus from Tables~\ref{tab:mem1} and
\ref{tab:mem2} and the subset of those candidates with full disks.
Extinction ranging from $A_K=0.2$--2.5 is displayed with the gray scale
\citep{juv16}. 
}
\label{fig:map56}
\end{figure}

As in \citet{luh10}, I have defined the disk fraction as the ratio of the
number of class II objects to the number of stars in classes II and III.
I also count evolved transitional disks as class~III.
Among the Lupus candidates in Tables~\ref{tab:mem1} and \ref{tab:mem2} that
have mid-IR data, the disk fraction is 82/131 ($0.63\pm0.04$) for on-cloud
stars and 13/62 ($0.21^{+0.06}_{-0.04}$) for off-cloud stars.
The lower disk fraction for the off-cloud stars is consistent with the
older ages that they exhibit in the CMDs (Section~\ref{sec:ages}).
The off-cloud stars are roughly coeval with members of Upper Sco in the
CMDs, and their disk fractions are similar as well \citep{luh20},
which supports the suggestion from Section~\ref{sec:ages} that most of
the off-cloud stars are members of Upper Sco.
To further illustrate that the youngest candidates are concentrated
near the Lupus clouds, I have plotted in Figure~\ref{fig:map56}
one map with all candidates from Tables~\ref{tab:mem1} and \ref{tab:mem2} and
a second map with only the candidates with the least evolved disks (full disks).

\subsection{Initial Mass Function}
\label{sec:imf}

\begin{figure}
\epsscale{1.2}
\plotone{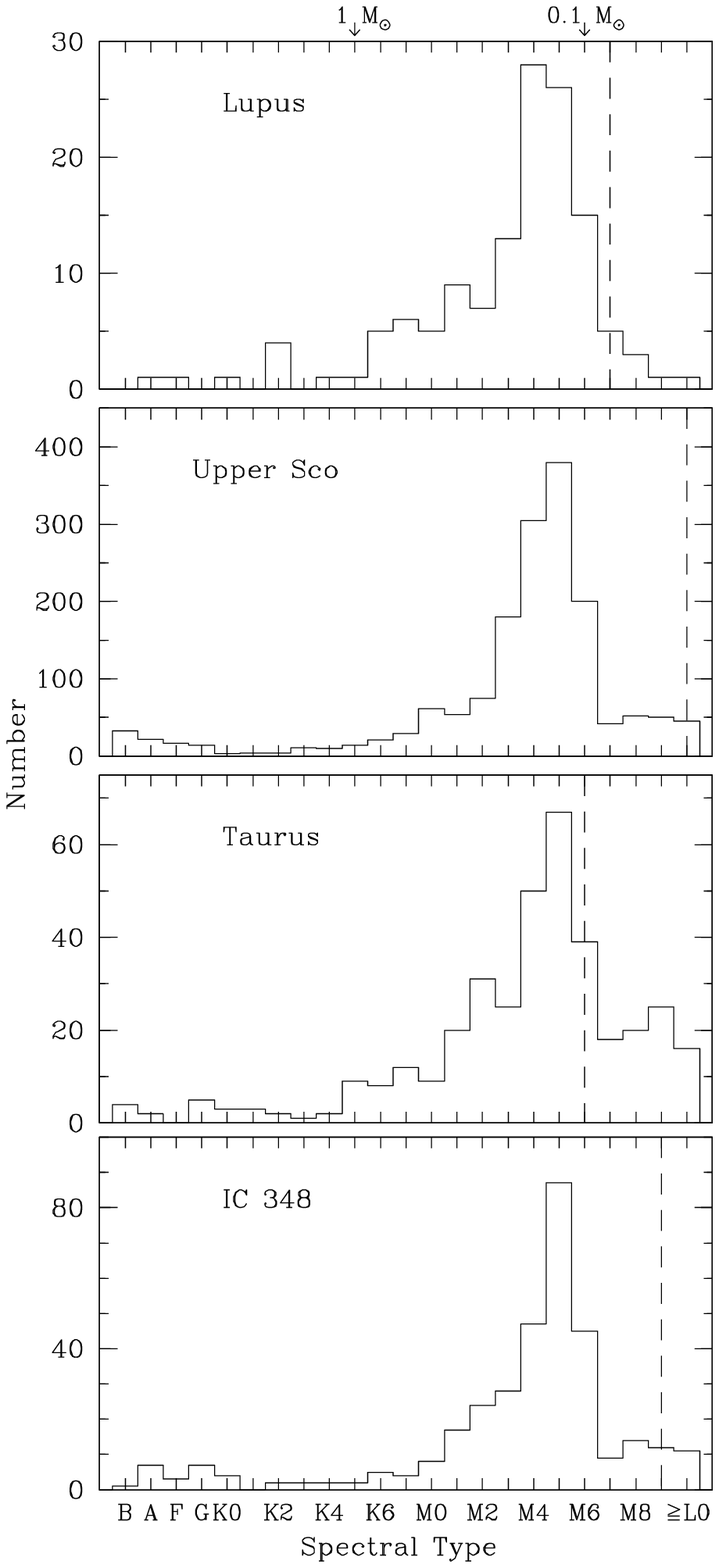}
\caption{
Distribution of spectral types for candidate members of Lupus from 
Tables~\ref{tab:mem1} and \ref{tab:mem2} that have $A_K<0.2$ and are
within the circular and elliptical fields in Figure~\ref{fig:map1}
and distributions for members of Upper Sco \citep{luh20}, Taurus
\citep[$A_J<1$,][]{esp19}, and IC~348 \citep[$A_J<1.5$,][]{luh16}.
The dashed lines indicate the completeness limits of these samples
and the arrows mark the spectral types that correspond to masses of 0.1
and 1~$M_\odot$ for ages of a few Myr according to evolutionary models
\citep[e.g.,][]{bar98,bar15}.
}
\label{fig:imf}
\end{figure}

As done in my previous work on nearby star-forming regions 
\citep[e.g.,][]{luh16,luh20}, I have characterized the IMF in Lupus
in terms of the distribution of spectral types, and I have constructed
that distribution for an extinction-limited sample of candidate members.
I have selected a limit of $A_K<0.2$ for that sample, which is high
enough to include a large majority of candidates while low enough that
the sample is complete down to low masses.
In Section~\ref{sec:apply}, I found that my survey of Lupus should be
complete down to $H\sim14$ for the on-cloud fields, which corresponds to
$\sim$M7 for $A_K=0.2$.
For candidates that lack spectral classifications, I have used spectral types
estimated from photometry in the manner described in Section~\ref{sec:spt}.
The top diagram in Figure~\ref{fig:imf} shows the distribution of spectral
types for on-cloud candidates from Tables~\ref{tab:mem1} and \ref{tab:mem2}
that have $A_K<0.2$. For comparison, I have included
distributions for samples of stars in Upper Sco \citep{luh20}, Taurus
\citep{esp19}, and IC~348 \citep{luh16}. The distributions for Lupus and 
those regions are similar, indicating similar IMFs.

\subsection{Radial Velocities and $UVW$ Velocities}
\label{sec:uvw}

Table~\ref{tab:mem1} includes previous measurements of radial velocities 
that have errors less than 4~km~s$^{-1}$, which are available for 78 Lupus
candidates. I have adopted errors of 0.4 and 1~km~s$^{-1}$ for velocities from
\citet{tor06} and \citet{wic99} for which errors were not reported,
respectively, which are near the typical precisions estimated in those studies.
For the 78 stars with radial velocities and Gaia measurements of proper motions
and parallaxes, I have used the radial velocities,
proper motions, and parallactic distances \citep{bai18} to compute $UVW$
space velocities \citep{joh87}, which are listed in Table~\ref{tab:mem1}.
Errors in the space velocities were estimated in the manner described
by \citet{luh20}.

Table~\ref{tab:uvw} lists the medians and standard deviations of the
$UVW$ velocities of candidates within the boundaries of clouds 1--4
from Figure~\ref{fig:map1}. These parameters are provided for each
cloud and for the combined sample of 69 stars from the four clouds.
The median velocities vary by a few km~s$^{-1}$ among the Lupus clouds
and they differ by a similar amount from the value of
($-5.1, -16.0, -7.2$)~km~s$^{-1}$ for Upper Sco \citep{luh20}.
The radial velocity errors tend to be greater than the equivalent errors
in proper motion, and the former contribute primarily to the errors in $U$,
which is why the standard deviations are largest in that velocity component.
The standard deviations of the velocities serve as upper limits on the
velocity dispersions.

\subsection{Constraints on Star Formation in Clouds 5 and 6}
\label{sec:clouds56}

Clouds 1--4 have been previously known to harbor star formation
based on the clustering of young stars near them on the sky
and the presence of significant reddening toward some of those stars
\citep[][references therein]{com08}. Clear evidence of star formation
has been lacking for the remaining Lupus clouds.
Two of those clouds, 5 and 6, are within the field selected for
my survey (Figure~\ref{fig:map1}).  The best available
constraints on star formation in clouds 5 and 6 have been provided
by Spitzer images \citep{spe11}, which were capable of identifying
disk-bearing stars at substellar masses and high extinctions.
\citet{spe11} found 15 candidates for disk-bearing stars based on
mid-IR excesses. Using parallaxes from Gaia DR2, \citet{man18} classified
eight candidates as background stars and four candidates as possible
members of clouds 5 and 6.  However, as mentioned in
Section~\ref{sec:compare}, I find that only one of
the latter stars has kinematics consistent with Lupus membership.
Given that the off-cloud Lupus candidates appear to be dominated by
Upper Sco members (Section~\ref{sec:ages}), that candidate is probably
a member of Upper Sco as well. The three remaining disk-bearing candidates
from \citet{spe11} lack Gaia parallaxes and confirmation of their youth.
Thus, it is likely that clouds 5 and 6 have not have experienced star formation.

\section{Conclusions}

I have performed a survey for stars associated with the Lupus clouds
using high-precision photometry and astrometry from Gaia DR2.
The new catalog of candidate members has been used to characterize the
age, disk fraction, IMF, and space velocities of the stellar population
of clouds 1--4 and to constrain the presence of star
formation in clouds 5 and 6. The results are summarized as follows:

\begin{enumerate}

\item
For an area within Sco-Cen that encompasses most of the Lupus clouds 
($l=335$--$345\arcdeg$, $b=5$--$19\arcdeg$), I have 1) identified candidate
young low-mass stars based on their positions in CMDs, 2) fit their kinematics
with a Gaussian mixture model that contains components for Lupus,
the V1062~Sco group, and UCL, 3) defined kinematic criteria for
membership in Lupus based on the stars in the Lupus component that
are grouped near clouds 1--4, 4) searched Gaia DR2 for stars within
the survey field that satisfy those criteria, and 5) rejected the Gaia
sources that appear to be too old for membership in Sco-Cen based on CMDs.
This process has produced 178 candidate members of Lupus.
I also have attempted to identify Lupus members that have Gaia kinematics 
but that failed my criteria due to erroneous astrometry, resulting in 13
additional candidates. These 178 and 13 candidates with Gaia kinematics
are presented in Table~\ref{tab:mem1}.

\item
I have examined previous IR surveys for viable Lupus candidates
near clouds 1--4 that lack measurements of kinematics from Gaia.
The resulting sample of 23 candidates is provided in Table~\ref{tab:mem2}.
The sample of candidates near clouds 1--4 from Tables~\ref{tab:mem1}
and \ref{tab:mem2} should be complete for spectral types earlier than
M7 at $A_K<0.2$.

\item
Based on the kinematic data from Gaia, many of the young stars previously
identified as possible members of Lupus are instead members of other
populations in Sco-Cen that overlap with the Lupus clouds (e.g., UCL).

\item
According to CMDs and disk fractions, the Lupus candidates far from
clouds 1--4 are older than the candidates near those clouds
and are coeval with the Upper Sco association (10--12~Myr).
Given that the off-cloud candidates (like all of the Lupus candidates) share
similar kinematics with Upper Sco, they are probably members of a low-density,
distributed component of Upper Sco that extends into the survey field.
Thus, the 144 on-cloud candidates in Tables~\ref{tab:mem1} and \ref{tab:mem2}
represent the best available sample of stars associated with the Lupus clouds.

\item
At spectral types of $\sim$K4--M5 in Gaia CMDs, the sequence of on-cloud Lupus
candidates is $\sim0.4$~mag brighter than the sequence for Upper Sco,
which implies an age of $\sim6$~Myr for Lupus assuming an age of 10--12~Myr
for Upper Sco and the change in luminosity with age predicted by
evolutionary models \citep{bar15,cho16,dot16,fei16}.

\item
I have used mid-IR photometry from WISE and the Spitzer Space Telescope
to check the Lupus candidates for excess emission from circumstellar disks
and have classified the evolutionary stages of the detected disks.
Most of the disks among these candidates have been reported in previous
studies, but I point out a few examples of new disks in more
advanced stages of evolution. The disk fraction for the on-cloud candidates
is N(II)/N(II+III)=$0.63\pm0.04$.

\item
Using spectroscopic and photometric estimates of spectral types,
I have constructed a distribution of spectral types for an
extinction-limited sample of on-cloud candidates. It is similar to the
distributions in other nearby star-forming regions, indicating a similar IMF.

\item
By combining Gaia parallaxes and proper motions with previous measurements of
radial velocities, I have calculated $UVW$ space velocities for 78 candidate
members of Lupus, including 69 of the on-cloud candidates. The median 
space velocities of the groups near clouds 1--4 vary by a few km~s$^{-1}$
and differ by a similar amount from the median velocity of Upper Sco.

\item
Previous Spitzer imaging has identified 15 candidate disk-bearing stars
toward clouds 5 and 6 \citep{spe11}, eight of which are background
stars based on Gaia parallaxes and three of which lack Gaia parallaxes
\citep{man18}. I find that the remaining four stars are probably members
of other groups in Sco-Cen. Thus, I find no evidence of stars associated
with clouds 5 and 6.

\end{enumerate}

\acknowledgements

I thank Taran Esplin and Eric Mamajek for comments on the manuscript.
This work used data from the European Space Agency (ESA)
mission Gaia (\url{https://www.cosmos.esa.int/gaia}), processed by
the Gaia Data Processing and Analysis Consortium (DPAC,
\url{https://www.cosmos.esa.int/web/gaia/dpac/consortium}). Funding
for the DPAC has been provided by national institutions, in particular
the institutions participating in the Gaia Multilateral Agreement.
2MASS is a joint project of the University of Massachusetts and IPAC
at Caltech, funded by NASA and the NSF. 
WISE is a joint project of the University of California, Los Angeles,
and the JPL/Caltech, funded by NASA.
This work used data from the Spitzer Space Telescope and the
NASA/IPAC Infrared Science Archive, operated by JPL under contract
with NASA, and the SIMBAD database, operated at CDS, Strasbourg, France.
The Center for Exoplanets and Habitable Worlds is supported by the
Pennsylvania State University, the Eberly College of Science, and the
Pennsylvania Space Grant Consortium.

\clearpage

\LongTables

\begin{deluxetable}{ll}
\tabletypesize{\scriptsize}
\tablewidth{0pt}
\tablecaption{Candidate Members of Lupus at $l=335$--$345\arcdeg$ and
$b=5$--$19\arcdeg$ that have Gaia Kinematics\label{tab:mem1}}
\tablehead{
\colhead{Column Label} &
\colhead{Description}}
\startdata
Gaia & Gaia DR2 source name \\
2MASS & 2MASS Point Source Catalog source name \\
WISEA & AllWISE Source Catalog source name \\
Name & Other source name \\
RAdeg & Right Ascension from Gaia DR2 (J2000) \\
DEdeg & Declination from Gaia DR2 (J2000) \\
SpType & Spectral type \\
r\_SpType & Spectral type reference\tablenotemark{a} \\
pmRA & Proper motion in right ascension from Gaia DR2\\
e\_pmRA & Error in pmRA \\
pmDec & Proper motion in declination from Gaia DR2\\
e\_pmDec & Error in pmDec \\
plx & Parallax from Gaia DR2\\
e\_plx & Error in plx \\
RVel & Radial velocity \\
e\_RVel & Error in RVel \\
r\_RVel & Radial velocity reference\tablenotemark{b} \\
U & $U$ component of space velocity \\
e\_U & Error in U \\
V & $V$ component of space velocity \\
e\_V & Error in V \\
W & $W$ component of space velocity \\
e\_W & Error in W \\
Gmag & $G$ magnitude from Gaia DR2\\
e\_Gmag & Error in Gmag \\
GBPmag & $G_{\rm BP}$ magnitude from Gaia DR2\\
e\_GBPmag & Error in GBPmag \\
GRPmag & $G_{\rm RP}$ magnitude from Gaia DR2\\
e\_GRPmag & Error in GRPmag \\
RUWE & renormalized unit weight error from \citet{lin18} \\
Jmag & $J$ magnitude \\
e\_Jmag & Error in Jmag \\
Hmag & $H$ magnitude \\
e\_Hmag & Error in Hmag \\
Ksmag & $K_s$ magnitude \\
e\_Ksmag & Error in Ksmag \\
JHKref & $JHK_s$ reference\tablenotemark{c} \\
W1mag & WISE W1 magnitude \\
e\_W1mag & Error in W1mag \\
f\_W1mag & Flag on W1mag\tablenotemark{d} \\
W2mag & WISE W2 magnitude \\
e\_W2mag & Error in W2mag \\
f\_W2mag & Flag on W2mag\tablenotemark{d} \\
W3mag & WISE W3 magnitude \\
e\_W3mag & Error in W3mag \\
f\_W3mag & Flag on W3mag\tablenotemark{d} \\
W4mag & WISE W4 magnitude \\
e\_W4mag & Error in W4mag \\
f\_W4mag & Flag on W4mag\tablenotemark{d} \\
ExcW2 & Excess present in W2? \\
ExcW3 & Excess present in W3? \\
ExcW4 & Excess present in W4? \\
Exc4.5 & Excess present in [4.5]? \\
Exc8.0 & Excess present in [8.0]? \\
Exc24 & Excess present in [24]? \\
DiskType & Disk Type \\
OnCloud & Near Lupus clouds 1--4?\tablenotemark{e}
\enddata
\tablecomments{This table is available in its entirety in a machine-readable form.}
\tablenotetext{a}{
(1) \citet{kra97};
(2) \citet{tor06};
(3) \citet{app83};
(4) \citet{hey89};
(5) \citet{hug94};
(6) \citet{koe00};
(7) \citet{her14};
(8) \citet{alc17};
(9) \citet{alc14};
(10) \citet{her77};
(11) \citet{alc20};
(12) \citet{mar94};
(13) \citet{hou78};
(14) \citet{pec16};
(15) \citet{mor11};
(16) \citet{rom12};
(17) \citet{can21};
(18) \citet{hou82};
(19) \citet{com13};
(20) \citet{muz14};
(21) \citet{all07};
(22) \citet{com03};
(23) \citet{man13};
(24) \citet{muz15};
(25) \citet{blo06};
(26) \citet{man14};
(27) \citet{mam02};
(28) \citet{nes95}.}
\tablenotetext{b}{
(1) Gaia DR2;
(2) \citet{fra17};
(3) \citet{gal13};
(4) \citet{tor06};
(5) \citet{alc20};
(6) \citet{gon06};
(7) \citet{wic99}.}
\tablenotetext{c}{
2 = 2MASS Point Source Catalog;
6 = 2MASS 6X Point Source Working Database \citep{cut12};
a = \citet{alc14};
v = sixth data release of the Visible and Infrared Survey Telescope for
Astronomy (VISTA) Hemisphere Survey \citep[VHS,][]{mcm13}.}
\tablenotetext{d}{nodet = non-detection; sat = saturated;
bl = photometry may be affected by blending with a nearby star;
bin = includes an unresolved binary companion; unres = too close to a brighter
star to be detected; false = detection from WISE catalog
appears false or unreliable based on visual inspection;
err = W2 magnitudes brighter than $\sim$6 are erroneous.}
\tablenotetext{e}{
Indicates whether the star is located within one of the circular and
elliptical fields encompassing clouds 1--4 in Figure~\ref{fig:map1}.}
\end{deluxetable}

\clearpage

\begin{deluxetable}{ll}
\tabletypesize{\scriptsize}
\tablewidth{0pt}
\tablecaption{On-cloud Candidate Members of Lupus that Lack Gaia Kinematics\label{tab:mem2}}
\tablehead{
\colhead{Column Label} &
\colhead{Description}}
\startdata
Gaia & Gaia DR2 source name \\
2MASS & 2MASS Point Source Catalog source name \\
WISEA & AllWISE Source Catalog source name \\
Name & Other source name \\
RAdeg & Right Ascension (J2000)\tablenotemark{a} \\
DEdeg & Declination (J2000)\tablenotemark{a} \\
SpType & Spectral type \\
r\_SpType & Spectral type reference\tablenotemark{b} \\
Gmag & $G$ magnitude from Gaia DR2\\
e\_Gmag & Error in Gmag \\
GBPmag & $G_{\rm BP}$ magnitude from Gaia DR2\\
e\_GBPmag & Error in GBPmag \\
GRPmag & $G_{\rm RP}$ magnitude from Gaia DR2\\
e\_GRPmag & Error in GRPmag \\
RUWE & renormalized unit weight error from \citet{lin18} \\
Jmag & $J$ magnitude \\
e\_Jmag & Error in Jmag \\
Hmag & $H$ magnitude \\
e\_Hmag & Error in Hmag \\
Ksmag & $K_s$ magnitude \\
e\_Ksmag & Error in Ksmag \\
JHKref & $JHK_s$ reference\tablenotemark{c} \\
W1mag & WISE W1 magnitude \\
e\_W1mag & Error in W1mag \\
f\_W1mag & Flag on W1mag\tablenotemark{d} \\
W2mag & WISE W2 magnitude \\
e\_W2mag & Error in W2mag \\
f\_W2mag & Flag on W2mag\tablenotemark{d} \\
W3mag & WISE W3 magnitude \\
e\_W3mag & Error in W3mag \\
f\_W3mag & Flag on W3mag\tablenotemark{d} \\
W4mag & WISE W4 magnitude \\
e\_W4mag & Error in W4mag \\
f\_W4mag & Flag on W4mag\tablenotemark{d} \\
ExcW2 & Excess present in W2? \\
ExcW3 & Excess present in W3? \\
ExcW4 & Excess present in W4? \\
Exc4.5 & Excess present in [4.5]? \\
Exc8.0 & Excess present in [8.0]? \\
Exc24 & Excess present in [24]? \\
DiskType & Disk Type 
\enddata
\tablecomments{This table is available in its entirety in a machine-readable
form.}
\tablenotetext{a}{
Right ascension and declination are from Gaia DR2, the 2MASS Point Source
Catalog, and the AllWISE Source Catalog in order of preference.}
\tablenotetext{b}{
(1) \citet{hey89};
(2) \citet{hug94};
(3) \citet{alc14};
(4) \citet{mor11};
(5) \citet{alc17};
(6) \citet{kra97};
(7) \citet{rom12};
(8) \citet{com13}.}
\tablenotetext{c}{
2 = 2MASS Point Source Catalog;
6 = 2MASS 6X Point Source Working Database \citep{cut12};
v = sixth data release of VISTA VHS.}
\tablenotetext{d}{nodet = non-detection; 
bl = photometry may be affected by blending with a nearby star;
ext = photometry is contaminated by extended emission;
bin = includes an unresolved binary companion; unres = too close to a brighter
star to be detected; false = detection from WISE catalog
appears false or unreliable based on visual inspection;
err = W2 magnitudes brighter than $\sim$6 are erroneous.}
\end{deluxetable}

\clearpage

\begin{deluxetable}{lccccccr}
\tabletypesize{\scriptsize}
\tablewidth{0pt}
\tablecaption{Medians and Standard Deviations of Space Velocities of
On-cloud Lupus Candidates\label{tab:uvw}}
\tablehead{
\colhead{Cloud} &
\colhead{$U$} &
\colhead{$V$} &
\colhead{$W$} &
\colhead{$\sigma_U$} &
\colhead{$\sigma_V$} &
\colhead{$\sigma_W$} &
\colhead{$N_*$}\\
\colhead{} &
\colhead{} &
\colhead{(km~s$^{-1}$)} &
\colhead{} &
\colhead{} &
\colhead{(km~s$^{-1}$)} &
\colhead{} &
\colhead{}}
\startdata
1 & $-$5.0 & $-$18.3 & $-$5.7 & 3.4 & 1.6 & 1.3 & 12\\
2 & $-$5.3 & $-$17.6 & $-$7.1 & 2.2 & 0.9 & 0.4 & 5\\
3 & $-$2.9 & $-$17.9 & $-$7.6 & 3.7 & 1.6 & 1.0 & 44 \\
4 & $-$4.1 & $-$18.1 & $-$7.3 & 1.8 & 0.9 & 0.8 & 8 \\
1--4 & $-$3.5 & $-$18.0 & $-$7.2 & 3.4 & 1.5 & 1.2 & 69
\enddata
\end{deluxetable}

\clearpage

\end{document}